\documentclass[journal=ancac3,doi=true]{achemso}
\usepackage{amssymb}
\usepackage{amsmath}
\usepackage{bbold}
\usepackage{mathrsfs} 
\usepackage{graphicx}
\usepackage{bm}
\usepackage{tabularx}
\usepackage[x11names]{xcolor}
\usepackage[unicode=true,colorlinks=false,hidelinks=true]{hyperref} 

\usepackage[normalem]{ulem}

\newcommand{\braket}[3]{\left\langle #1 \left| #2 \right| #3 \right\rangle}

\title{Confined acoustic phonons in CsPbI\textsubscript{3} nanocrystals explored by resonant Raman scattering on excitons}
\makeatletter \let\thetitle\@title \makeatother

\author{Carolin Harkort$^{1}$}
\author{Ina V. Kalitukha$^{1}$}
\author{Nataliia E. Kopteva$^{1}$} 
\author{Mikhail O. Nestoklon$^{1}$} 
\email{nestoklon@gmail.com}
\author{Serguei V. Goupalov$^2$}
\author{Lucien Saviot$^3$} 
\author{Dennis Kudlacik$^{1}$}
\author{Dmitri R. Yakovlev$^{1}$}
\email{dmitri.yakovlev@tu-dortmund.de}
\author{Elena V. Kolobkova$^{4,5}$}
\author{Maria S. Kuznetsova$^{6}$}
\author{Manfred~Bayer$^{1,7}$}

\affiliation{
  $^{1}$Experimentelle Physik 2, Technische Universit\"at Dortmund, 44227 Dortmund, Germany\\
  $^{2}$Department of Physics, Jackson State University, Jackson, Mississippi 39217, USA\\
  $^{3}$Universit\'e Bourgogne Europe, CNRS, Laboratoire Interdisciplinaire Carnot de Bourgogne ICB UMR 6303, 21000 Dijon, France\\
  $^{4}$ITMO University, 199034 St. Petersburg, Russia\\
  $^{5}$St. Petersburg State Institute of Technology, 190013 St. Petersburg, Russia\\
  $^{6}$Spin Optics Laboratory, St. Petersburg State University, 198504 St. Petersburg, Russia\\
  $^{7}$Research Center FEMS, Technische Universit\"at Dortmund, 44227 Dortmund, Germany
}
\makeatletter \let\theauthors\acs@author@list \let\theaffillist\acs@address@list \makeatother

\begin{document}

\begin{tocentry}
\includegraphics{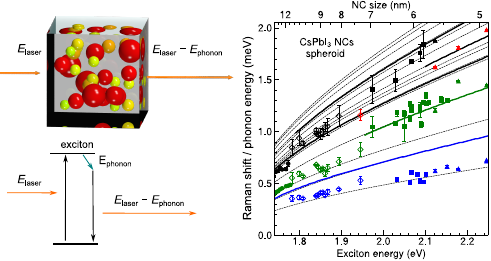}
\end{tocentry}

\begin{abstract}



Optical properties of  the lead halide perovskites nanocrystals are controlled by confined  excitons and rich spectrum of  confined acoustic and optical phonons. We study experimentally and theoretically the exciton-phonon interaction in CsPbI$_3$ perovskite nanocrystals embedded in a glass matrix. Energies of phonon modes allowed by selection rules are detected by resonant Raman scattering for nanocrystals with sizes of 4–13~nm, covering exciton energies of 1.72–2.25~eV. While optical phonon energies remain size-independent, the energies of confined acoustic phonons increase  in smaller nanocrystals. Acoustic phonons are modeled within the continuum approximation using elastic constants computed by density functional theory. The model provides the energy spectra of confined phonons for nanocrystals of various shapes (cube, sphere, spheroid), crystal symmetries (orthorhombic and tetragonal), and sizes. Exciton confinement restricts efficient coupling to only few phonon modes observable in Raman spectra. By comparing experimental data with model predictions, we conclude that the nanocrystals in our samples predominantly have spherical or spheroidal shapes.
\end{abstract}

\maketitle


\section{}
 
Lead halide perovskite semiconductors have gained significant attention due to their exceptional photovoltaic efficiency~\cite{jen1019,nrel2021}, potential for optoelectronic~\cite{Vinattieri2021_book,Vardeny2022_book}  and spintronic applications~\cite{Vardeny2022_book,wang2019,ning2020,kim2021}. Perovskite nanocrystals (NCs) have recently been included in the family of semiconductor colloidal NCs that can be synthesized in solution~\cite{Kovalenko2017,Chen2018a,Protesescu2015,Akkerman2018,Park2015,Yu2021}. They exhibit a quantum yield of up to 90\%, even for unpassivated NCs, as surface states do not significantly impair the exciton emission efficiency. The structural symmetry of perovskite NCs can be manipulated through temperature and composition adjustments, while the synthesis technique primarily determines their shape and size. Consequently, perovskite nanocrystals are an extended testbed for exploring the relationships between crystal symmetry, NC shape and size, as well as their influence on optical and elastic properties.

Perovskite NCs are commonly synthesized by colloidal chemistry in solution and can be composed of hybrid organic-inorganic or fully-inorganic materials. The fully-inorganic NCs made, e.g., of CsPbI$_3$, CsPbBr$_3$, or CsPbCl$_3$, show considerably higher stability in ambient conditions compared with the materials containing organic components. Encapsulating the nanocrystals further improves their stability. One effective approach to achieve this is the synthesis of perovskite nanocrystals within a glass matrix directly from a melt.~\cite{Li2017,Liu2018,Liu2018a,Ye2019,kolobkova2021,Belykh2022}. The surface of such nanocrystals is free from organic ligands, which are typically present in solution-grown NCs. Additionally, glass-embedded samples can be easily polished to obtain optically smooth surfaces, making them well-suited for optoelectronic applications.

Nanocrystals (quantum dots) embedded in a glass matrix were historically the first semiconductor system in which quantum confinement of electrons and holes was demonstrated~\cite{ekimov1981,efros1982,ekimov1985}. Later the focus shifted towards colloidal nanocrystals grown in solution~\cite{bawendi1993,Talapin2010} due to the simpler and more controllable technology and to the emergence of Stranski-Krastanov quantum dots \cite{marzin1994,alferov1995} which allow for easier device integration. 

While the most characteristic property of semiconductor NCs is the quantum confinement of charge carriers, their nanometer size also leads to the quantization of acoustic phonons, which can be observed by different methods \cite{ng2022}, including low-frequency Raman scattering. Such quantization has been demonstrated for semiconductor NCs, including Si~\cite{fujii1996}, CdSe~\cite{saviot1996}, CdS~\cite{sirenko1998,saviot1998}, PbS~\cite{Krauss1997}, and PbSe~\cite{Ikezawa2001}.   Recently, it has been shown that the size of nanodiamonds can be determined from shifts in Raman lines~\cite{vlk2022}.
The electron-phonon interaction in halide perovskites is of great interest due to strong polaron effects \cite{Miyata2017}, which in particular strongly affect their transport properties \cite{Ponce2019}. In studies of the optical properties of nanocrystals, most attention has been focused on optical phonons  \cite{Fu2017,Cho2021,Iary2021,Amara2023}, while experimental observations of acoustic phonons confined within the nanocrystals are rare~\cite{Lv2021}.

In this work, we reveal how confined acoustic phonons and low-energy optical phonons affect resonantly excited excitons in Raman scattering spectroscopy of CsPbI$_3$ nanocrystals embedded in a glass matrix. Four samples with different nanocrystal sizes, covering an exciton energy range of 1.72--2.25~eV (diameters of 4--13~nm), are examined. We employ elastic constants for CsPbI$_3$ obtained via density functional theory (DFT) calculations and apply the continuum approximation to compute the energies of confined acoustic phonon modes and to model the Raman scattering spectra. In the calculations, various structural symmetries (orthorhombic and tetragonal), shapes (cube, sphere, spheroid), and sizes are taken into account. By comparing the theoretical results with experimental data, we conclude that the shapes of the CsPbI$_3$ nanocrystals in our samples are close to spheres or spheroids.



\begin{figure*}[t]
\begin{center}
\includegraphics[width=\textwidth]{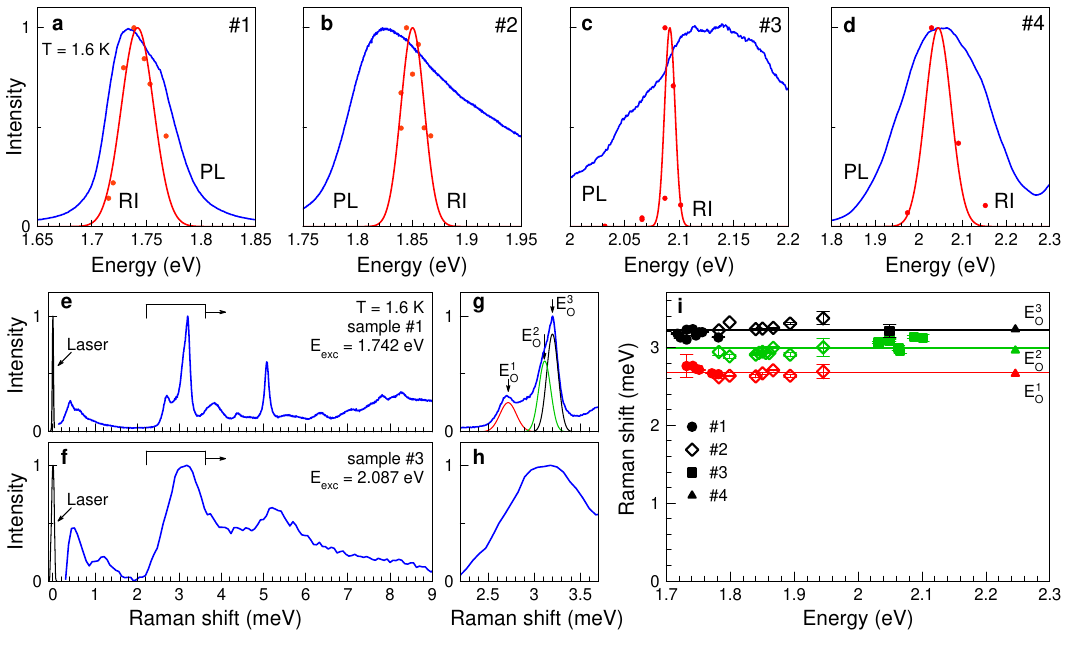}
\caption{Optical properties of CsPbI$_3$ NCs at $T =1.6~$K. Photoluminescence (blue) spectra for samples: \#1 (a), \#2 (b), \#3 (c), and \#4 (d). The photon excitation energy for the PL measurements is $3.49~$eV. The red symbols show the Raman intensity as function of the excitation energy. The red lines are guides for the eye. Spectrum of phonons active in Raman scattering in samples \#1 (e) and \#3 (f). Laser power density is $0.06$~W/cm$^2$ and $E_\text{exc} = 1.742~$eV for (e).  Power density is $0.3$~W/cm$^2$ and $E_\text{exc} = 2.087$~eV for (f). Detection and excitation are co linearly polarized.  (g,h) Close-up of spectra for Raman shift energies from 2.2~meV to 3.7~meV. $E_\text{O}^1$, $E_\text{O}^2$, and $E_\text{O}^3$ denote the optical phonon shifts, extracted by fits using three Gaussian functions. (i) Dependence of the optical phonon shifts on excitation energy.
\label{fig:PL}  }
\end{center}
\end{figure*}

We experimentally study a series of CsPbI$_3$ nanocrystals embedded in a fluorophosphate glass matrix. The samples, labeled as \#1, \#2, \#3, and \#4, vary in mean NC diameters from 13 to 4~nm, respectively. Figure~\ref{fig:PL}a-d shows the photoluminescence (PL) spectra measured at the temperature of $T = 1.6$~K. The PL lines are inhomogeneously broadened with a full width at half maximum ranging from 120~meV to 250~meV, due to the significant dispersion in NC size within each sample.

We use resonant excitation with a spectrally-narrow continuous-wave laser to address the exciton properties corresponding to a particular NC size. The intensity of the Raman scattering, measured across the exciton energy range of $1.72-2.25$~eV, is shown in Figure~\ref{fig:PL}a-d by the symbols. The spin coherence signal measured by a pump-probe method in these samples also has a narrower spectral range than the PL spectrum~\cite{meliakov2025}. Detailed information on the optical and magnetooptical properties, as well as the carrier Land\'e $g$-factors of the studied NCs can be found in Refs.~\citenum{nestoklon2023,meliakov2024,meliakov2025}.



Figure~\ref{fig:PL}e presents a typical Raman spectrum measured on sample \#1 at $T =1.6$~K. A positive Raman shift corresponds to the value of the Stokes shift, i.e., the signal is shifted to lower energy from the laser line. The Raman spectrum shows many spectral lines in a range covering 9~meV of Raman shifts. The multiple spectral lines with energies larger than 2~meV correspond to exciton-mediated Raman scattering on optical phonons. The Raman shifts of these lines are 2.70, 3.10, 3.20, 3.80, 4.40, 5.08, 5.67, 6.36, 7.05, 7.80, and 8.26\,meV. Figure~\ref{fig:PL}g shows the zoomed spectrum of the three lowest optical phonon lines, labeled as $E_\text{O}^1$, $E_\text{O}^2$, and $E_\text{O}^3$. For Raman shifts below 2~meV, lines corresponding to confined acoustic phonons are seen. 

Examples of Raman spectra of sample \#3 with smaller NCs, which are measured at larger photon energy of 2.087~eV, are shown in Figures~\ref{fig:PL}f,h. One can see that the Raman lines for optical and acoustic phonons broaden, but the optical phonons do not change their Raman shifts compared to sample \#1. We measured the phonon energies in the excitation range from 1.72~eV to 2.25~eV in the four samples, probing nanocrystals with different sizes (4--13~nm). As one can see in Figure~\ref{fig:PL}i, the Raman shifts of the optical phonon modes remain independent of the excitation energy (i.e. independent of the NC size), as is expected for the optical phonons. Experimental and theoretical investigations of the optical phonons in CsPbBr$_3$ NCs can be found in Ref.~\citenum{bataev2024}.

\begin{figure*}[t]
\begin{center}
\includegraphics[width =\textwidth]{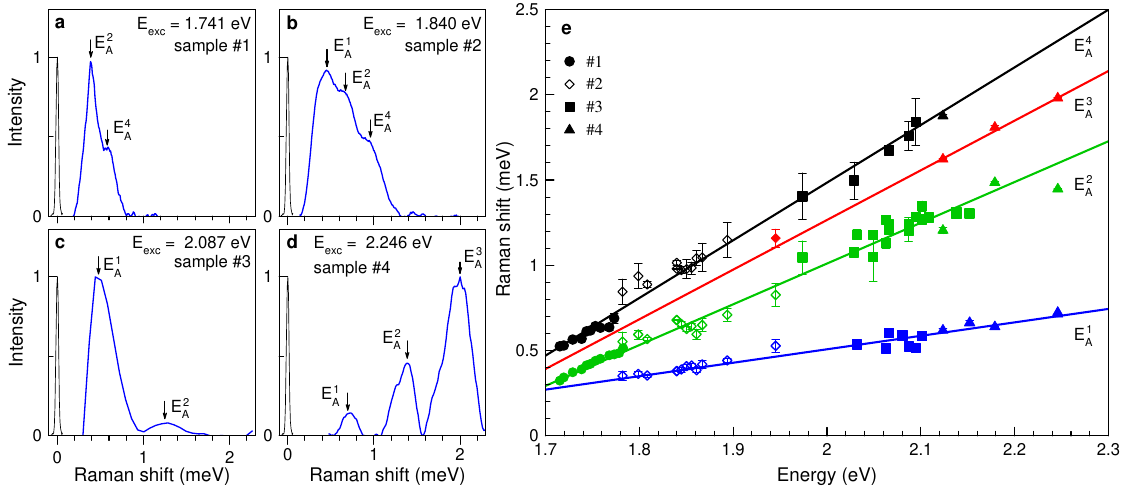}
\caption{Raman spectra of acoustic phonons for CsPbI$_3$ NCs of different sizes at $T = 1.6$~K. Spectra are measured at $E_\text{exc} = 1.741$~eV on sample \#1 (a), at $E_\text{exc} = 1.840$~eV on sample \#2 (b), at $E_\text{exc} = 2.087$~eV on sample \#3 (c), and at $E_\text{exc} = 2.246$~eV on sample \#4 (d). The energies of the acoustic phonons are labeled as $E^i_\text{A}$, with $i = 1,2,3,4$ in each spectrum. (e) Acoustic phonon energies as function of excitation energy for the four samples. Colors indicate the specific phonon modes: blue ($E^1_\text{A}$), green ($E^2_\text{A}$), red ($E^3_\text{A}$), and black ($E^4_\text{A}$). Symbols denote data from different samples: circles (sample \#1), open diamonds (\#2), squares (\#3), and triangles (\#4). Lines are guides to the eye. 
\label{fig:spec_ph}}
\end{center}
\end{figure*}

Let us focus on the acoustic phonon peaks at Raman shifts smaller than 2~meV, see Figure~\ref{fig:spec_ph}a. The arrows mark the two pronounced peaks at energies $E_\text{A}^2 = 0.40$~meV and $E_\text{A}^4 = 0.59$~meV, measured at the excitation energy of $E_\text{exc} = 1.742$~eV. The Raman scattering intensity in the Stokes range is higher than in the anti-Stokes range, as shown in Figure~\ref{fig:pol_dep} in the Supporting Information~\ref{sec:Raman_si}. 
Although acoustic phonon energies are identical in both ranges,
not all modes are clearly resolved in the anti-Stokes component.
A longitudinal magnetic field ($B$) does not modify the acoustic phonon lines, as shown in Figure~\ref{fig:ph_MF}a in Supporting Information.

 The Raman shifts of the acoustic phonon lines increase for larger excitation energies, i.e. for smaller NCs, as shown in Figures~\ref{fig:spec_ph}a-d for samples with different nanocrystal sizes. We fit the lines using Gaussian functions and plot their Raman shifts ($E_\text{A}^{i}$) as functions of the excitation energy in Figure~\ref{fig:spec_ph}e.  Raman shifts of confined acoustic phonons increase with the excitation energy in range of $1.72$~eV to $2.25$~eV. Using linear extrapolations, the $E_\text{A}^1$ mode changes its energy from 0.25 to 0.75~meV and the $E_\text{A}^4$ mode from 0.45 to 2.50~meV. Understanding the size dependence of confined acoustic phonon energies requires the theoretical analysis and modeling presented below.


The Raman scattering in nanocrystals (NCs) can be modeled as a three-step process involving exciton generation by the incident light, exciton–phonon interaction, and exciton recombination accompanied by photon emission \cite{Martin1971}. 
For the qualitative analysis we consider spherical nanocrystals composed of an isotropic material, where the crystal anisotropy and the anisotropic part of the deformation potential are neglected. In spherical NCs, only interaction with the spheroidal confined phonon modes with total angular momentum $\ell = 0$ and $\ell = 2$ is allowed by symmetry \cite{duval1992}. For lead halide perovskites, which have a simple band structure and isotropic deformation potentials, only the $\ell = 0$ modes contribute significantly to the Raman signal \cite{goupalov1999}.

At the first step, we calculate the amplitude of the confined phonon mode inside the NC \cite{goupalov1999}: 
\begin{equation}\label{eq:GM_I2} \begin{split} I_2(x) = \frac{x^4}{\gamma_l \left[\zeta(x) + x(x \cos x - \sin x)\right]^2 + x^2 \zeta^2(x)}, \\ \zeta(x) = 4 \gamma_{tl}(\rho_{io} - \gamma_t) \left(\cos x - \frac{\sin x}{x} \right) + \rho_{io} x \sin x. \end{split} 
\end{equation}
Here, the velocity contrast parameters are defined as
$\gamma_l = (c_l^{o} / c_l^{i})^2$,
$\gamma_{tl} = (c_t^{o} / c_l^{o})^2$,
$\gamma_t = (c_t^{o} / c_t^{i})^2$,
where the subscripts $l$ and $t$ denote the longitudinal and transverse sound velocities, and the superscripts $i$ and $o$ refer to the values inside and outside the nanocrystal, respectively. The density ratio is given by $\rho_{io} = \rho_i / \rho_o$. 

\begin{figure}
  \centering{\includegraphics[width=0.5\textwidth]{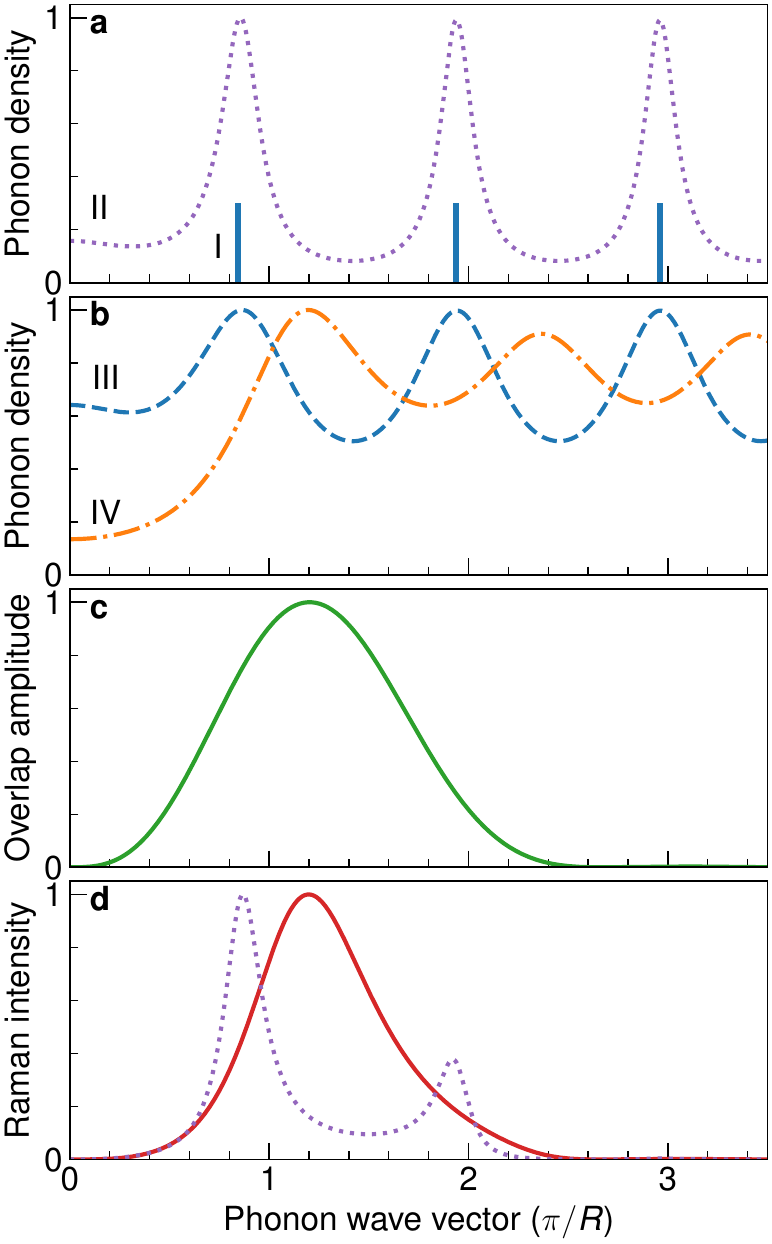}}
  \caption{(a) The dotted line shows the phonon amplitudes in the nanocrystal, calculated using Eq.~\eqref{eq:GM_I2} with a high density contrast but no velocity contrast (parameters Set II from Table~\ref{tbl:clt}). The vertical lines indicate the discrete phonon wave numbers confined in the NC, calculated using free boundary conditions (Set I).
(b) Phonon amplitudes in NCs calculated with Set III (realistic density contrast without velocity contrast, dashed line) and Set IV (realistic contrast of both density and velocity of the matrix, dash-dotted line).
(c) Overlap of phonon and exciton envelope functions, calculated using Eq.~\eqref{eq:GM_I1}.
(d) Raman signal amplitude as function of the characteristic phonon wave vector $q$, calculated using Eq.~\eqref{eq:GM} with Sets IV (solid line) and II (dotted line).
}
\label{fig:GMsband}
\end{figure} 

A free-standing nanocrystal with free boundary conditions has discrete phonon modes depicted as vertical lines in Figure~\ref{fig:GMsband}a. Paramerets for calculation are given in Table~\ref{tbl:clt} (Set I). If nanocrystal is in glass matrix with a density five times smaller, then the phonon modes broaden, although their energies remain mostly unchanged (dotted line in Figure~\ref{fig:GMsband}a; Set II from Table~\ref{tbl:clt}). The broadening can be qualitatively attributed to the leakage of the displacement field into the surrounding matrix.

\begin{table}\caption{Sound velocities used in the calculations. Densities are given in g/cm$^3$ and velocities in $10^{5}$~cm/s.}\label{tbl:clt}
\begin{tabular*}{\linewidth}{@{\extracolsep{\fill}}lrrrrrr}
 \hline
 \hline
 & $\rho^{i}$ & $c_l^{i}$ & $c_t^{i}$  & $\rho^{o}$ & $c_l^{o}$ & $c_t^{o}$  \\
 \hline
  I  & $5.2$& $2.4$ & $1.3$ & ---   & --- & ---\\
 II  & $5.2$& $2.4$ & $1.3$ & $1$   & $2.4$ & $1.3$\\
III  & $5.2$& $2.4$ & $1.3$ & $3.7$ & $2.4$ & $1.3$\\
IV   & $5.2$& $2.4$ & $1.3$ & $3.7$ & $4.0$ & $2.4$\\
 \hline
 \hline
 \end{tabular*}
 \end{table}

To compute the Raman spectra, we consider the exciton–phonon overlap function $I_1(x)$: 
\begin{equation}\label{eq:GM_I1} I_1(x) = x^3 \left[ \int_{0}^{1} \sin^2(\pi t) \frac{\sin (xt)}{xt} \, \mathrm{d}t \right]^2, \quad x = qR, 
\end{equation}
as a function of the phonon wave vector $q$, shown in Figure~\ref{fig:GMsband}c. It selects phonon modes with wave vectors comparable to quantum-confined electrons or holes wave vectors. 
The Raman intensity for a phonon mode with the wave vector $q$ in a NC of radius $R$ is given by \cite{goupalov1999,goupalov2000}: \begin{equation}\label{eq:GM} I(q;R) \sim \frac{1}{R^3} I_1(qR) I_2(qR). \end{equation} 
The resulting Raman spectrum consisting of the phonon peaks selected by the envelope function is shown in Figure~\ref{fig:GMsband}d.
Using realistic glass density (Set III) enhances the effect of the surrounding matrix, and the phonon peaks become broadened without shifting their positions (dashed line in Figure~\ref{fig:GMsband}b). Adding a sound velocity contrast (Set IV) further shifts the phonon energies to higher values due to the velocity mismatch, while the broadening remains similar (dash-dotted line in Figure~\ref{fig:GMsband}b; solid line in Figure~\ref{fig:GMsband}d).


For CsPbI$_3$, the simple model of spherical nanocrystals made of isotropic material is not applicable, since the bulk crystal exhibits strong elastic anisotropy due to its orthorhombic symmetry. Therefore, we account for both material anisotropy and nanocrystal shape in our phonon quantization analysis: 
We compute the acoustic phonon states numerically (for details, see Refs.~\citenum{saviot2009,saviot2021}) and simulate the Raman spectra using Eq.~\eqref{eq:GM_I1}. Elastic constants are calculated by density functional theory (DFT), with the PBEsol exchange-correlation potential \cite{Perdew2008}, as implemented in the WIEN2k code \cite{Blaha2020} (see Supporting Information~\ref{sec:DFT} and Table~\ref{tbl:Cij}).

Our calculations yield a set of phonon eigenfrequencies $\Omega_i$ and corresponding volume variations: 
\begin{equation}
\label{eq:dV}
\delta V_i V^{2/3} = \left| \int_{V} \bm{\nabla} \cdot {\bf u}_i({\bf r}) , \mathrm{d}^3{\bf r} \right|, 
\end{equation} 
where ${\bf u}_i({\bf r})$ is the displacement field of the $i$-th phonon mode at position ${\bf r}$, as defined in Ref.~\citenum{saviot2009}. The integral runs over the NC volume $V$.

Since the exciton-phonon interaction in perovskites mainly stems from the isotropic deformation potential and proportional to $\bm{\nabla} \cdot {\bf u}({\bf r})$ (see the Supporting Information~\ref{sec:GM}), for the Raman spectrum calculations we replace Eq.~\eqref{eq:GM} with
\begin{equation}\label{eq:Inum}
  I(\Omega;R) \sim \frac{I_1\left({\Omega R}/{\left\langle c \right\rangle} \right)}{R}  \; \sum_{i} \delta V_i \tilde{\delta}(\Omega-\Omega_i)\,.
\end{equation}
Here $\left\langle c \right\rangle = 1.4\times 10^5$~cm/s is the average sound velocity (see Ref.~\citenum{Anderson1963}) and the Gaussian broadening of the modes is $\tilde{\delta}(x) = \exp\left[-x^2/\sigma^2 \right]$ with the parameter $\sigma=30$~$\mu$eV. The quantity $\hbar \Omega$ corresponds to the Raman shift. This expression captures the full phonon quantization and an approximate exciton–phonon interaction.

\begin{figure*}
  \centering{\includegraphics[width=\textwidth]{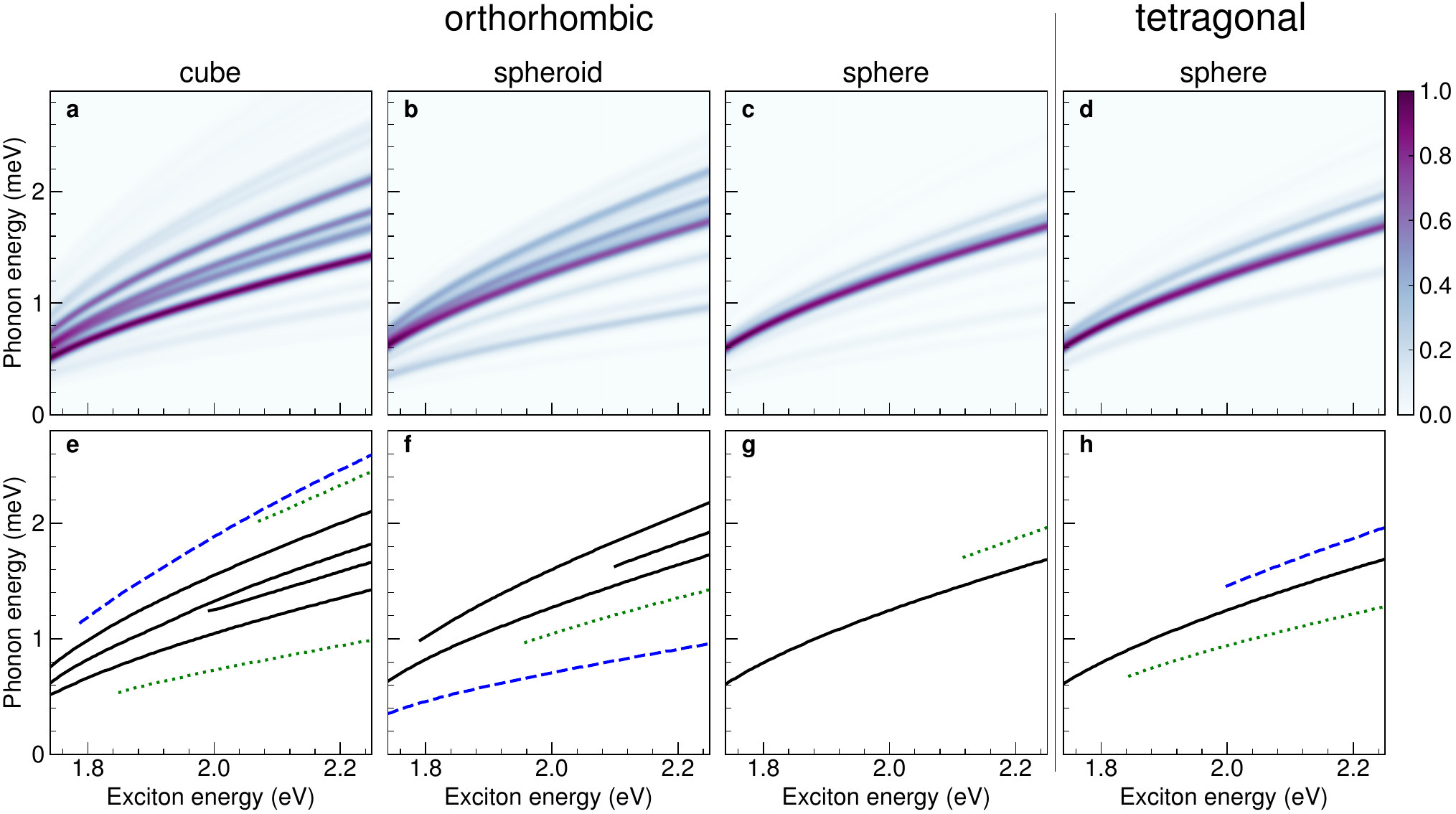}}
  \caption{(a-d) Amplitude and shift of the Raman lines as function of the exciton energy reflecting the NC size. Calculations are made according to Eq.~\eqref{eq:Inum} for: (a) cubic, (b) spheroidal (with aspect ratio $D_{x,z}/D_{y} = 1.8$), and (c) spherical CsPbI$_3$ NCs with orthorhombic crystal structure. (d) Results for spherical NCs with tetragonal structure. (e-h) Raman shifts of the strong lines taken from panels (a-d). 
Peaks with larger intensity are shown by the solid lines, with intermediate intensity by the dashed blue lines, and with small intensity by the dotted green lines.
}
\label{fig:theory}
\end{figure*} 

Figure~\ref{fig:theory}a-d shows the Raman active confined phonon modes as functions of the resonant excitation energies depending on the NC size. For each NC size, the phonon mode energies are computed numerically, and the Raman intensity, represented by the color scale, is estimated using the interpolation of the calculations from Ref.~\citenum{nestoklon2023}; (see Supporting Information~\ref{sec:NC_size}). Figure~\ref{fig:theory}e-h display the phonon energies to highlight weaker Raman lines. Our calculations consider NCs of different shapes: spheres, spheroids, and cubes. We also compare the spherical case's elastic parameters for CsPbI$_3$ in both the tetragonal and orthorhombic phases. All presented simulations  assume free boundary conditions.

Figure~\ref{fig:theory}  shows that the average Raman peak energy primarily depends on the NC size (inferred from the exciton energy), while the number of distinct peaks varies with the NC shape. In spherical NCs, a single strong peak is expected. In spheroids, an additional peak appears due to the activation of the spheroidal $\ell = 2$ mode. In cubic NCs, the splitting of the low-energy peak is so complex that it is completely smeared out. Instead, higher-energy phonon modes dominate in the Raman spectra. The richness of the calculated spectra arises from the absence of rotational symmetry and the simple selection rules for isotropic spheres \cite{duval1992} no longer apply.
The calculated spectra shown in Figure~\ref{fig:theory} depend sensitively on the NC shape, which can be determined from the Raman spectra. This approach is a promising analytic tool, as NCs in glass are not easy to analyze. For example, transmission electron microscopy (TEM) is complicated by charge accumulation.

\begin{figure}[t]
  \centering{\includegraphics[width=0.6\textwidth]{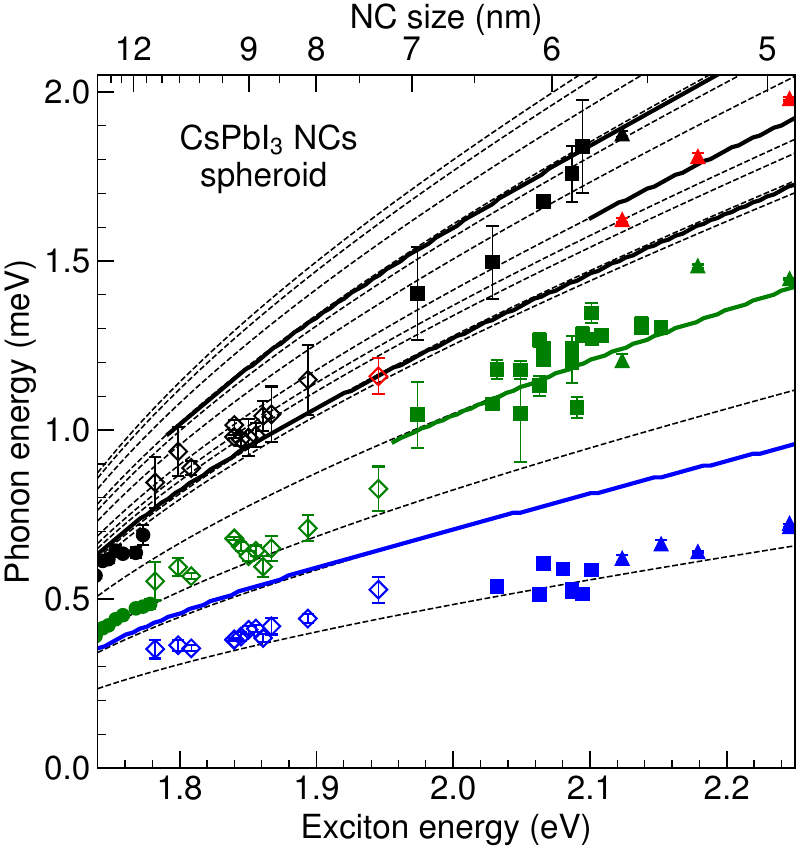}}
  \caption{Lowest 20 phonon modes with symmetry $A_g$ in the spheroidal CsPbI$_3$ NCs calculated as function of exciton energy. Dashed lines show the positions of all phonon modes and thicker color lines reproduce the calculated Raman active peaks from Figure~\ref{fig:theory}f. Experimental data are shown by symbols. 
  }
\label{fig:theory_all}
\end{figure}

In Figure~\ref{fig:theory_all} we compare the measured energies of the acoustic phonons with theoretical calculations for spheroid NCs, which give the best agreement with experiment. Thick lines trace strongest modes in the Raman spectra and dashed lines show other phonon modes with the $A_g$ symmetry \cite{Wherrett1986}. 
Note, the higher energy peak observed experimentally agrees well with the energy of the calculated phonon mode originating from the fundamental breathing mode of the isotropic sphere. The lowest Raman peak is in good agreement with the lowest energy phonon mode, which originates from the fundamental spheroidal phonon mode with  $\ell=2$ (see Supporting Information~\ref{sec:GM}). The calculated Raman spectra and their comparison with experimental spectra and the mechanisms responsible for the amplitude mismatch are discussed in Supporting Information~\ref{sec:GM}.

The detailed comparison of the calculated and experimental Raman spectra are given in Fig.~\ref{fig:theory_spectra}.  From the comparison it is clear that the position and shape of the Raman peaks are unambiguously connected with the NC size and shape. Therefore, resonant Raman scattering is a very valuable technique for the structural characterization of perovskite NCs.

The strong interaction between Raman-active confined acoustic phonons and excitons makes these phonons a key factor influencing exciton emission properties as well as their coherent and incoherent dynamics. For example, an exciton can relax from optically allowed bright states to an optically forbidden dark state by emitting one or several phonons. Therefore, a deeper understanding of the properties of confined acoustic phonons is essential for understanding the optical properties of halide perovskite nanocrystals, since the energy scale of confined phonon modes is on the same order of magnitude as the exciton fine structure splitting \cite{Fu2017,Nestoklon2018,Tamarat2022,Han2022}.


In conclusion, lead halide perovskite nanocrystals (NCs) exhibit a rich spectrum of confined acoustic phonon modes. Some of these modes are optically active and can be detected using resonant Raman scattering. We experimentally investigated these phonons in CsPbI$_3$ nanocrystals, ranging in size from 4 to 13 nm, embedded in a glass matrix. The energies of the confined phonons are between 0.5 and 2.0 meV and increase as the size of the nanocrystals decreases. Our theoretical analysis includes calculation of the elastic tensor for bulk CsPbI$_3$ in both tetragonal and orthorhombic phases. We also modeled the phonon spectra for nanocrystals of various shapes (cubic, spherical, and spheroidal) and simulated the Raman spectra by considering the interaction between Raman active confined phonons and excitons. By comparing the calculated Raman spectra with experimental data, we conclude that the nanocrystals are predominantly spherical or spheroidal in shape. This work demonstrates that resonant Raman spectroscopy, when combined with theoretical modeling, is a powerful optical method for probing the symmetry, shape, and size of perovskite nanocrystals. Additionally, this approach can be used to study surface boundary conditions and the interaction between nanocrystals and their surrounding environment. A comparative study of nanocrystals embedded in a glass matrix versus those grown in solution holds excellent potential for further insights. 

\section{Methods}
\label{sec:Methods}

\textbf{Samples.} The studied CsPbI$_3$ nanocrystals embedded in fluorophosphate Ba(PO$_3$)$_2$-AlF$_3$ glass were synthesized by rapid cooling of a glass melt enriched with the components needed for the perovskite crystallization. Details of the method are given in Ref.~\citenum{kolobkova2021}. Samples of fluorophosphate (FP) glass with the composition 35P$_2$O$_5$–35BaO–5AlF$_3$–10Ga$_2$O$_3$–10PbF$_2$–5Cs$_2$O (mol. \%) doped with BaI$_2$ were synthesized using the melt-quench technique. The glass synthesis was performed in a closed glassy carbon crucible at the temperature of $T = 1050^\circ\text{C}$. Four samples are investigated in this paper, which we label \#1, \#2, \#3, and \#4. Their technology codes are EK8, EK201, EK5, and EK205, respectively. About 50~g (for sample \#1) and 25~g (for samples \#2, \#3, and \#4) of the batch were melted in the crucible for $20–30$~minutes, and then the glass melt was cast on a glassy carbon plate and pressed to form a plate with a thickness of about 2~mm. Samples with a diameter of 5~cm were annealed at the temperature of 50$^\circ\text{C}$ below $T_g = 400^\circ\text{C}$ to remove residual stresses. The CsPbI$_3$ perovskite NCs were formed from the glass melt during the quenching. The glasses obtained in this way are doped with CsPbI$_3$ NCs. The NC sizes in the initial glass were regulated by the concentration of iodide and the melt's cooling rate without heat treatment above $T_g$. They differ in the NC sizes, which is reflected by the relative spectral shifts of their optical spectra. The change in the NC size was achieved by changing the concentration of iodine in the melt. Due to the high volatility of iodine compounds and the low viscosity of the glass-forming fluorophosphate melt at elevated temperatures, an increase in the synthesis time leads to a gradual decrease of the iodine concentration in the equilibrium melt. Thus, it is possible to completely preserve the original composition and change only the concentration of iodine employing a smooth change of the synthesis duration.\\
\textbf{Magneto-optical measurements.} The samples were placed in a helium bath cryostat with a variable temperature insert for low-temperature optical measurements. At $T = 1.6$~K the sample is placed in superfluid helium. A superconducting magnet with split coils generates a magnetic field of up to 10 Tesla. The magnetic field was applied parallel to the light vector (Faraday geometry).\\
\textbf{Photoluminescence and absorption measurements.} The photoluminescence spectrum was measured with an 0.5~m spectrometer equipped with a charge-coupled-devices (CCD) camera. The sample was illuminated from the back side with the halogen lamp for the absorption measurement.\\
\textbf{Raman scattering.} The Raman scattering signal was excited by a single-frequency laser tunable from 2.31~eV to 2.45~eV (Syrah company laser system based on a Ti:Sapphire laser with the spectral range extended by a frequency doubling unit and by mixing with the light of a fiber laser emitting at 1950~nm). 
The laser was focused on the sample to a spot with a diameter of about 300~$\mu$m, the laser power was set to 0.2~mW for sample \#1,  corresponding to an excitation density of $0.06$~W/cm$^2$, and 1~mW for samples \#2, \#3, and \#4, corresponding to $0.3$~W/cm$^2$. 
The scattered light was analyzed by a Jobin-Yvon U1000 double monochromator with 1-meter focal length, allowing the high resolution of 0.2~cm$^{-1}$ (0.024~meV). The Raman signal was detected by a cooled GaAs photomultiplier and conventional photon-counting electronics. The spectrally narrow laser, high spectral resolution of the spectrometer, and efficient suppression of the scattered laser light allow us to detect Raman shifts from 0.1 to 3~meV. The Raman spectra were measured for co-polarized linear polarizations of excitation and detection. 


\subsection*{Data Availability Statement}
The data presented in this paper are available from the corresponding authors upon reasonable request.

\subsection*{Acknowledgements}
We acknowledge the financial support by the Deutsche Forschungsgemeinschaft: C.R. and D.R.Y. (via the SPP2196 Priority Program, project YA65/28-1, no. 527080192), M.O.N (project AK40/13-1, no. 506623857),  N.E.K. (project KO 7298/1-1, no. 552699366), and I.V.K. (project KA 6253/1-1, no. 534406322). The work of S.V.G. was supported by the NSF through DMR-2100248.
We acknowledge the computing time provided on the Linux HPC cluster at TU Dortmund (LiDO3), partially funded in the course of the Large-Scale Equipment Initiative by the DFG as project 271512359. 
L.S. acknowledges support by the EIPHI Graduate School (contract ANR-17-EURE-0002) operated by the French National Research Agency (ANR). E.V.K. and M.S.K. acknowledge the Saint-Petersburg State University (Grant No. 125022803069-4).

\clearpage
\newpage
\onecolumn
\begin{center}
  \textsf{\textbf{\Large Supporting Information \\  \thetitle}} \\
  {\textsf{\theauthors}\par}
  {\textit{\theaffillist}\par}
\end{center}
\setcounter{equation}{0}
\setcounter{figure}{0}
\setcounter{table}{0}
\setcounter{page}{1}
\setcounter{section}{0}
\makeatletter
\renewcommand{\thepage}{S\arabic{page}}
\renewcommand{\theequation}{S\arabic{equation}}
\renewcommand{\thefigure}{S\arabic{figure}}
\renewcommand{\thetable}{S\arabic{table}}
\renewcommand{\thesection}{S\arabic{section}}
\renewcommand{\bibnumfmt}[1]{[S#1]}
\renewcommand{\citenumfont}[1]{S#1}
\makeatletter
\let\@startsection\acs@startsection
\makeatother


\section{Additional Raman spectra}
\label{sec:Raman_si}

\begin{figure*}[t]
\begin{center}
\includegraphics[width=\textwidth]{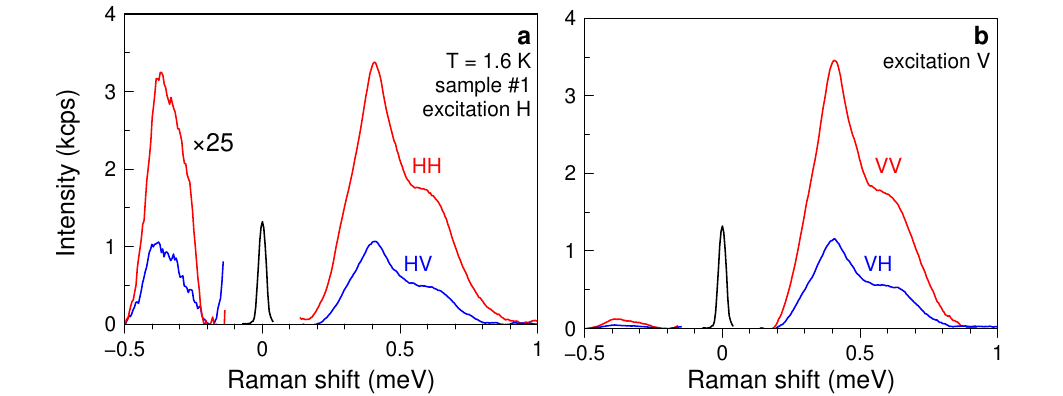}
\caption{Polarization properties of the Raman scattering spectra for sample \#1 measured at an excitation energy of $E_\text{exc} = 1.742$~eV at temperature $T = 1.6$~K. (a) Raman spectra detected in horizontal (red) and vertical (blue) polarizations under horizontally (H) polarized laser excitation. The HH configuration indicates co-polarized excitation and detection, both in the horizontal direction, while HV corresponds to crossed-polarized excitation (horizontal) and detection (vertical). The peaks in the anti-Stokes spectral region are magnified by a factor of 25 to enhance visibility. (b) Raman spectra recorded for horizontal (blue) and vertical (red) polarizations under vertically (V) polarized laser excitation. The VV configuration represents co-polarized excitation and detection in the vertical direction, while VH denotes crossed-polarized excitation (vertical) and detection (horizontal).
\label{fig:pol_dep}}
\end{center}
\end{figure*}

In Figure~\ref{fig:pol_dep} Raman spectra with Stokes (positive Raman shifts) and anti-Stokes (negative shifts) components are shown. The spectra are measured with parallel (HH, VV) and crossed (HV, VH) linear polarizations of the laser and detected signal. The polarization of the signal follows the polarization of the exciton peaks and one may conclude that the interaction with the phonons is not spin-dependent. 

By comparing the intensities of the Stokes ($I_{S}$) and anti-Stokes ($I_{aS}$) Raman signals one can estimate \cite{IvchenkoBooksi} the phonon temperature by means of 
\begin{equation}
  \frac{I_{S}}{I_{aS}} = \exp{\left(\frac{\Delta E}{kT}\right)} \,.
\end{equation}
In our experiment, $I_{S}/I_{aS} \approx 25$, which gives $T \approx 1.5$~K coinciding with the expected lattice temperature of the sample in contact with superfluid helium. We conclude that laser heating of the studied NCs is negligibly small in our experimental conditions. 

\section{Magnetic field dependence}
\label{sec:MF}

\begin{figure*}[t]
\begin{center}
  \includegraphics[width = 0.8\textwidth]{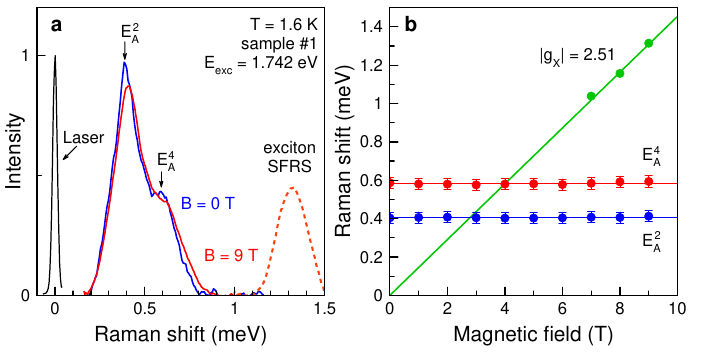}
\caption{Acoustic phonons in the Raman spectra measured on sample \#1 at $T = 1.6$~K. (a) Raman spectra at zero magnetic field (blue line) and in a longitudinal magnetic field of $B = 9$~T (red line). The dotted red line gives the exciton spin-flip Raman scattering (SFRS). $E_\text{exc} = 1.742~$eV. (b) The Raman shifts for the acoustic phonons $E^2_\text{A}$ and $E^4_\text{A}$ (red and blue) and the exciton spin-flip (green) in dependence on the longitudinal magnetic field. The symbol size of the spin-flip data includes the error bar. The line is a linear fit giving the exciton $g$-factor of $|g_\text{X}| = 2.51 \pm 0.01$.}
\label{fig:ph_MF}
\end{center}
\end{figure*}

A longitudinal magnetic field ($B$) does not modify the acoustic phonon lines, as shown in Figure~\ref{fig:ph_MF}a. The Raman shift of the lines $E_\text{A}^2$ and $E_\text{A}^4$ is independent of the field strength up to 9~T, see Figure~\ref{fig:ph_MF}b. An additional Raman line appears in magnetic fields exceeding 7~T (red dashed line in Figure~\ref{fig:ph_MF}a). Its Raman shift increases linearly with magnetic field, as it is equal to the Zeeman energy splitting $g_{\text{X}} \mu_{\text{B}} B$, with $\mu_\text{B}$ being the Bohr magneton. The fit gives a value of $|g_\text{X}| = 2.51\pm 0.01$, which is typical for the exciton Zeeman splitting in CsPbI$_3$ NCs~\cite{meliakov2025si}.

\section{DFT calculations}
\label{sec:DFT}

For the calculations of confined acoustic phonons in NCs, we need all nine elastic constants and material density for bulk CsPbI$_3$. The existing experimental data on the elastic constants of lead halide perovskites are limited. Therefore, we use the constants calculated using density functional theory (DFT). Note that the parameters are rather sensitive to the choice of the exchange-correlation potential, and the use of experimental data would be preferable, if/when available. 

For the calculations, we use the package WIEN2k \cite{Blaha2020si}. First, the structure is optimized. For the cubic phase, only the lattice constant is changed to have the minimum total energy. For the tetragonal phase, also the position of halide atoms in the $xy$ plane should be optimized. For the orthorhombic phase the positions of Cs and halide atoms should be relaxed. In the optimization of internal positions, the atoms are moved until the forces at all atoms are sufficiently small. Next, the components of the elastic tensor are calculated with the help of the IRelast package \cite{Jamal2018si}. In this package, a set of deformed structures is generated, and for each type of deformation, the energy as function of the deformation amplitude is calculated. Note that for low symmetry phases, the positions of the atoms within the primitive cell should also be optimized for each deformation value. From a quadratic fit of the calculated energy, combinations of elastic tensor components are extracted. The calculation of the optimal structure and the strain tensor components are rather sensitive to the precision settings. A detailed analysis shows that the ``automatic'' precision setting \verb|-prec 3n| with fine $k$ mesh (about two thousand points) is enough to obtain a precision better than 1\%. The energy is converged up to $10^{-5}$ a.u., for the forces in most cases the value $0.1$~a.u. is used. The results of the calculations are given in Table~\ref{tbl:Cij}. 

\begin{table*}\caption{Components of the elastic tensor for a bulk CsPbI$_3$ crystal calculated in the DFT approach for three phases: cubic $\alpha$-CsPbI$_3$, tetragonal $\beta$-CsPbI$_3$, and orthorhombic $\gamma$-CsPbI$_3$. Density $\rho$ in g/cm$^3$, $C_{ij}$ are in GPa. }
\label{tbl:Cij}
 \begin{tabular*}{\linewidth}{@{\extracolsep{\fill}}lrrrrrrrrrr}
 \hline
 \hline
                   & $\rho$ & $C_{11}$ & $C_{22}$ & $C_{33}$ & $C_{12}$  & $C_{13}$ & $C_{23}$ & $C_{44}$ &  $C_{55}$ & $C_{66}$\\
 \hline
$\alpha-$CsPbI$_3$ & $4.909$ & $42.9$ & $42.9$ & $42.9$ & $ 4.49$ & $4.49$ & $4.49$ & $3.89$ & $ 3.89$ & $3.89$\\
$\beta-$CsPbI$_3$  & $5.072$ & $26.9$ & $26.9$ & $45.6$ & $18.7$  & $18.7$ & $7.05$ & $5.51$ & $ 5.51$ & $16.1$ \\
$\gamma-$CsPbI$_3$ & $5.160$ & $25.3$ & $31.7$ & $27.9$ & $ 9.49$ & $13.1$ & $17.3$ & $6.86$ & $15.8$  & $6.62$ \\
 \hline
 \hline
 \end{tabular*}
 \end{table*}


%

\section{Modeling Raman spectra of excitons interacting with confined acoustic phonons in NC$\rm s$}
\label{sec:GM}


The Raman scattering probability is given by~\cite{LLIVsi}
\begin{equation}
  W(\hbar\omega_i,\hbar\omega_f) = \frac{2\pi}{\hbar} \sum_{if} \left| M_{fi} \right|^2 \delta(\hbar\omega_i,\hbar\omega_f\mp \hbar\Omega)\,,
\end{equation}
where 
\begin{equation}
  M_{fi} = \sum_{n,m} \frac{  \braket{0}{V^*}{n} \braket{n,N\pm1}{H_{\rm X-ph}}{m,N} \braket{m}{V}{0}}
  {(\hbar\omega_i-E_{X,m}-i\Gamma/2)(\hbar\omega_f-E_{X,n}-i\Gamma/2)} .
\end{equation}
Here $\braket{0}{V^*}{n}$ and $\braket{m}{V}{0}$ are the matrix elements of creation/recombination of an exciton in a NC, $\braket{n,N\pm1}{H_{X-ph}}{m,N}$ is the matrix element of the exciton-phonon interaction, $\hbar\omega_i$ and $\hbar\omega_j$ are the energies of the incident and emitted photons, respectively, $\hbar\Omega$ is the phonon energy. $E_{X,m}$ is the $m$-th exciton energy and $\Gamma$ is the exciton damping rate. $N$ is the number of phonons.

Assuming abrupt boundary conditions at the NC surface and a strong confinement regime (Bohr radius large as compared to NC size), the envelope of the exciton wave function is the product of the electron and hole envelopes:
\begin{equation}
  \Psi({\bf r}_e,{\bf r}_h) = \psi({\bf r}_e)\psi({\bf r}_h)\,,
\end{equation}
where ${\bf r}_{e(h)}$ is the electron (hole) coordinate and the envelopes of the fundamental states of both electron and hole in spherical perovskite NC with radius $R$ are: 
\begin{equation}
  \psi({\bf r}) = \frac1{\sqrt{2\pi R}} \frac{\sin(\pi r /R)}{r}\,.
\end{equation}


The interaction between exciton and phonons microscopically originates from the deformation potential Hamiltonian
\begin{equation}\label{eq:Hdef}
  H_{\rm X-ph} \equiv H_{\rm{def}} ({\bf r}) = a_X \bm{\nabla} \cdot {\bf u}\,,
\end{equation}
where ${\bf u}$ is the displacement due to the phonons and the exciton deformation potential $a_X=a_c-a_v$ is the difference between the deformation potentials in conduction and valence bands.

It is well known that only phonon modes with total angular momenta $\ell=0$ and $\ell=2$ are Raman active \cite{duval1992si,gupalov2006si}. The modes with $\ell=2$ need some microscopic mechanism that allows such transitions, and below, we consider only modes with zero total angular momentum. The displacement corresponding to the breathing mode and its overtones is:
\begin{equation}
  {\bf u}({\bf r}) = \frac{A_0}{\sqrt{2\pi}} j_{1}(qr) {\bf e_r}\,,
\end{equation}
where $A_0$ is the normalization coefficient and $j_1$ is the spherical Bessel function of first order. This means that for $\ell=0$
\begin{equation}
  H_{\rm{def}} ({\bf r}) = a_{X} A_0 \frac{\sin(qr)}{r}\,.
\end{equation}

Thus, the result for the exciton-phonon interaction is Eqs.~(\ref{eq:GM}-\ref{eq:GM_I2}) (Eq.~(38) from \cite{goupalov1999si}). When the real spectrum of an anisotropic material is considered, this result may be generalized to give Eq. \eqref{eq:Inum}. Let us briefly outline the limitations and approximations used in this approach. For a detailed microscopic theory, $I_1$ from Eq.~\eqref{eq:GM_I1} should be calculated using the actual wave functions of electron and hole bound in the exciton that is confined in a NC. Also, Eq.~\eqref{eq:Hdef} does not take into account distant bands, which might lead to an anisotropic exciton-phonon interaction.

\begin{figure*}
  \centering{\includegraphics[width=\textwidth]{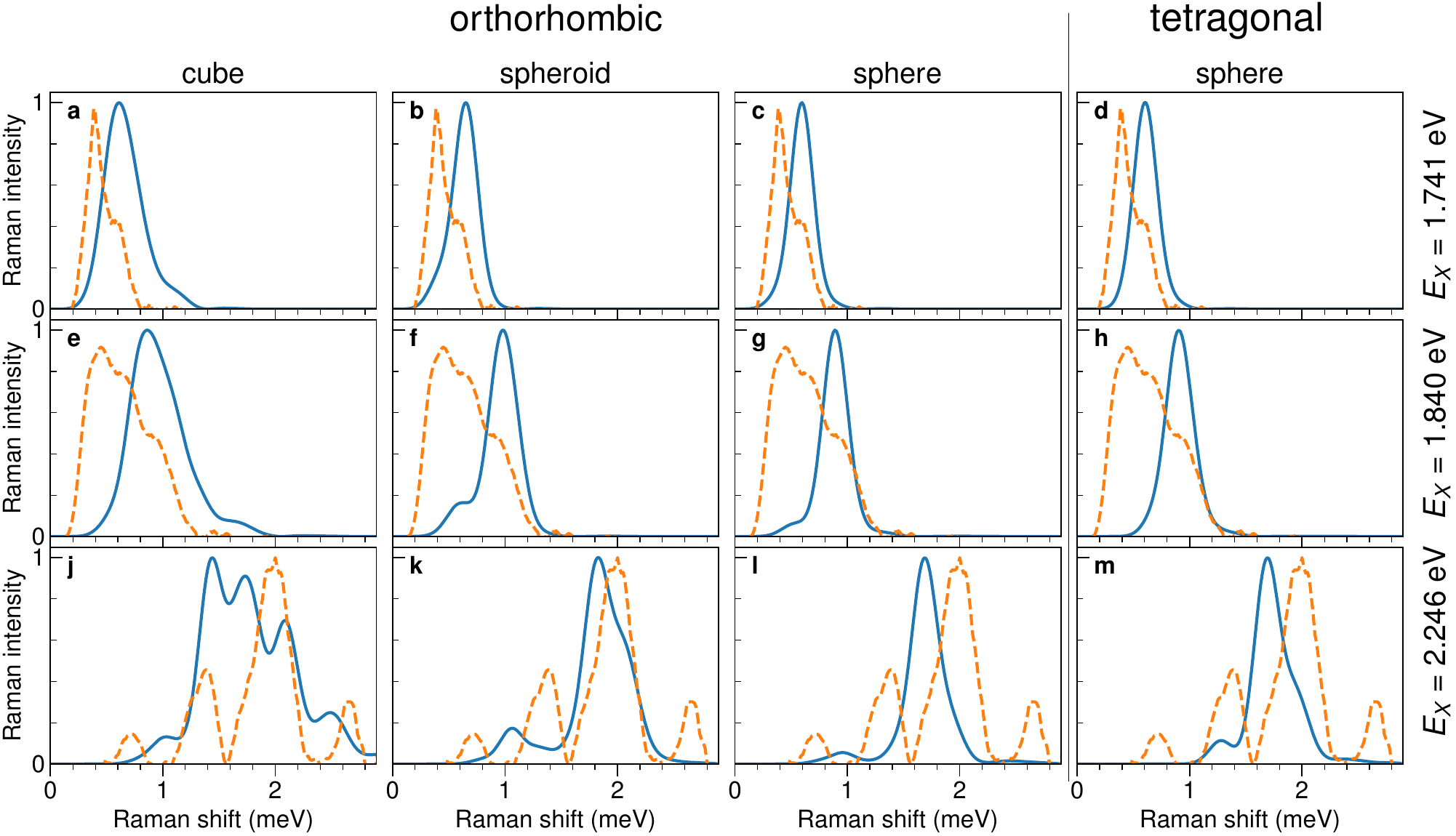}}
  \caption{Comparison of theoretical (lines) and experimental (dashed) Raman scattering spectra for three typical NC sizes. The rows correspond to different exciton peak energies of 1.741~eV (upper row), 1.840~eV (middle row), and 2.246~eV (lower row). The columns correspond to the spectra calculated for cubic (1st column), spheroidal (2nd column), spheric (3rd column) and spheric from tetragonal material (4th column). 
  }
\label{fig:theory_spectra}
\end{figure*}

In Fig.~\ref{fig:theory_spectra} we show the comparison of the calculated spectra with the experimental data for three exciton energies (for large, intermediate, and small NCs in the first, second, and third row, respectively). The calculations are done for cubic, spheroid, and spherical NC in the orthorhombic phase and for spherical NC in the tetragonal phase. The fine structure with  splitting comparable with the experimental data is clearly seen for the spheroid NC, while for the spherical NC this structure has only one distinct peak and for the cubic NC the expected result is a signal with the amplitude evenly distributed within the fine structure of phonon modes. Note, however, that the shape of the peaks for large NCs in the experimental data has a much stronger low-energy part, see the discussion in the main text. 

The calculated spectra shown in Figure~\ref{fig:theory} in the main text depend sensitively on the NC shape. For more details, we provide Raman spectra for NCs of different shapes and sizes in the Supporting Information Figure~\ref{fig:theory_spectra}. The difference is due to a change in the ``fine structure'' of the phonon modes. From this result, the NC shape can be determined from the Raman spectra. This approach is a promising analytic tool, as NCs in glass are not easy to analyze using other methods. In particular, transmission electron microscopy (TEM) is complicated by charge accumulation and problems with sample preparation.

However, there is a difference between the calculations and the experimental data. In Figure~\ref{fig:theory_all} by dashed lines we show the first 20 phonon modes with the $A_g$ symmetry \cite{Wherrett1986si}. The thick lines trace modes which are expected to be the strongest in the Raman spectra. They coincide with the modes shown in Figure~\ref{fig:theory}f and are encoded by the same colors.
One can see that the experimental peak positions have the same dependence on the NC size as these of the calculated Raman-active confined phonon modes. The higher energy peak observed experimentally agrees well with the energy of the calculated phonon mode originating from the fundamental breathing mode of the isotropic sphere. The lowest Raman peak is in good agreement with the lowest energy phonon mode, which originates from the fundamental spheroidal phonon mode with  $\ell=2$. Technically, the exciton-phonon interaction with this mode is allowed by symmetry already in the isotropic approximation \cite{duval1992si}. However, for simple electron/hole bands this interaction is zero \cite{goupalov1999si} and an additional mechanism, which lowers the symmetry, is required to make this mode  optically active. The non-zero amplitude of this mode can be explained either by its mixing with the $\ell=0$ mode by the anisotropy of the phonon mode (due to anisotropy of elastic parameters and/or NC shape) or by the anisotropic exciton-phonon interaction arising from the interaction with distant bands.
Other sources of the difference may be some surface effects and/or the role of the matrix in the exciton-phonon interaction neglected in our analysis.

Let us discuss the role of the boundary conditions.
From Figures~\ref{fig:GMsband}a,b in the main text, it is clear that the Raman signal also depends on the details  of the conditions imposed on the displacement field at the boundary between a NC and a glass matrix. In particular, the broadening of the Raman signal observed experimentally (see Figures~\ref{fig:ph_MF} and \ref{fig:spec_ph}) may not be governed by the experimental resolution or the NC size distribution, but rather be originated from the boundary conditions at the NC/glass matrix interface, if they are mechanically connected. However, if the mechanical contact at the surface is not ideal due to the technology of the sample preparation, then the phonon modes are localized fully within the NCs, and the broadening of the peaks caused by the phonon mode leakage to the glass matrix should be suppressed.
Additional studies are needed to determine which of the options is realized in our samples. Nevertheless, we note that Raman spectroscopy is the unique tool to investigate not only the NCs themselves, but also their interaction with the environment.

To conclude, some quantitative discrepancy in the measured and calculated energies of the phonon modes may  be explained by several reasons: (i) the limited accuracy of the elastic constants obtained from the DFT calculations, (ii) the limited accuracy of the experimental peak positions (the width of the peaks in the experimental Raman spectra is comparable or smaller than the distance between them), and (iii) the role of the boundary conditions. 



\section{Quantum confinement energy versus NC size}
\label{sec:NC_size}

To compare the calculated data with the experimental Raman spectra we need to associate the exciton energies $E_X$ with the NC size. For this purpose, we use the interpolation of the empirical tight-binding (ETB) calculations described in Ref.~\citenum{nestoklon2023si}. In this work, the energy is calculated for cubic NCs with an integer number of atomic layers. For the estimation of the NC size from the exciton peak energy we have separated the interpolation for cubic and sperical NCs. For cubic, the energy is given by 
\begin{equation}\label{eq:EX_interp_c}
  E_X = 1.652+\frac{11}{2.8+L^2} \,,
\end{equation}
where $L$ is the length of the NC edge, and for spherical NCs
\begin{equation}\label{eq:EX_interp_s}
  E_X = 1.652+\frac{16}{3.307+D^2} \,,
\end{equation}
where $D$ is the NC diameter. Note that Eqs.~(\ref{eq:EX_interp_c},\ref{eq:EX_interp_s}) are justified by the fact that the confinement energy for complex shaped NCs depends mostly on the NC volume so that a similar confinement energy is expected for a cubic NC with edge $L$ and a  spherical NC with diameter $D\approx 2L/{(4 \pi /3)^{1/3}}$.

\begin{figure}[t]
  \centering{\includegraphics[width=0.7\textwidth]{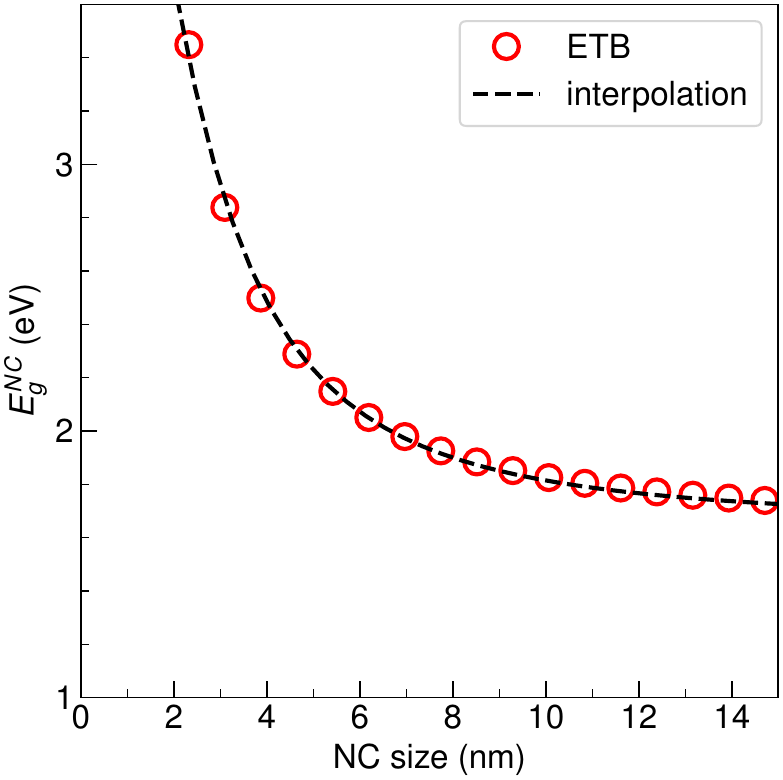}}
  \caption{Dependence of the optical transition energy on the NC size calculated for CsPbI$_3$ NCs.
The empirical tight-binding calculations are taken from Ref.~\citenum{nestoklon2023si}, the size of the NC is recalculated to an effective diameter of a spherical NC. For the interpolation we use Eq.~\eqref{eq:EX_interp_s}.
}
\label{fig:theory_size}
\end{figure}

In Figure~\ref{fig:theory_size} we show the comparison of the ETB results from Ref.~\citenum{nestoklon2023si} with Eq.~\eqref{eq:EX_interp_s}. Note that in the calculations \cite{nestoklon2023si} the exciton binding energy and its renormalization by the quantum confinement is neglected. The difference is expected~\cite{galkowski2016si} to be less than 20~meV, which may be safely neglected for the purpose of the current consideration.

\newpage
\begin{thebibliography}{99}

\bibitem{jen1019} Jena, A. K.; Kulkarni, A.; Miyasaka, T.
Halide perovskite photovoltaics: Background, status, and future prospects.
\textit{Chemical Reviews} 	\textbf{2019}, 5, {3036--3103}.

\bibitem{nrel2021} Best Research - Cell Efficiency Chart,
\url{https://www.nrel.gov/pv/cell-efficiency.html} (2025).

\bibitem{Vinattieri2021_book} \textit{Halide Perovskites for Photonics}; Vinattieri, A.; Giorgi, G., Eds.; AIP Publishing: Melville, New York, \textbf{2021}.

\bibitem{Vardeny2022_book} \textit{Hybrid Organic Inorganic Perovskites: Physical Properties and Applications}; Vardeny, Z. V; Beard, M. C., Eds.; World Scientific, \textbf{2022}.

\bibitem{wang2019} Wang, J.; Zhang, C.; Liu, H.; McLaughlin, R.; Zhai, Y.; Vardeny, S. R.; Liu, X.; McGill, S.; Semenov, D.; Guo, H.; Tsuchikawa, R.; Deshpande, V. V.; Sun, D.; Vardeny, Z. V.
Spin-optoelectronic devices based on hybrid organic-inorganic trihalide perovskites.
\textit{Nat. Commun.} 	\textbf{2019}, 10, {129}.

\bibitem{ning2020} Ning, W.; Bao, J.; Puttisong, Y.; Moro, F.; Kobera, L.; Shimono, S.; Wang, L.; Ji, F.; Cuartero, M.; Kawaguchi, S.; Abbrent, S.; Ishibashi, H.; De Marco, R.; Bouianova, I. A.; Crespo, G. A.; Kubota, Y.; Brus, J.; Chung, D. Y.; Sun, L.; Chen, W. M.;  Kanatzidis, M. G; Gao, F.
Magnetizing lead free halide double perovskites.
\textit{Science Advances} 	\textbf{2020}, 6, {eabb5381}.

\bibitem{kim2021} Kim, Y.-H.; Zhai, Y.; Lu, H.; Pan, X.; Xiao, C.; Gaulding, E. A.; Harvey, S. P.; Berry, J. J.; Vardeny, Z.~V.; Luther, J. M.; Beard, M. C.
Chiral-induced spin selectivity enables a room-temperature spin light-emitting diode.
\textit{Science} 	\textbf{2021}, 371, {1129}.

\bibitem{Kovalenko2017} Kovalenko, M. V.; Protesescu, L; Bondarchuk, M. I.
Properties and potential optoelectronic applications of lead halide perovskite nanocrystals.
\textit{Science} \textbf{2017}, 358, 745--750.

\bibitem{Chen2018a} Chen, Q.; Wu, J.; Ou, X.; Huang, B.; Almutlaq, J.; Zhumekenov, A. A.; Guan, X.; Han, S.; Liang, L.; Yi, Z.; Li, J.; Xie, X.; Wang, Y.; Li, Y.; Fan, D.; Teh, D. B. L.; All, A. H.; Mohammed, O. F.; Bakr, O. M.; Wu, T.; Bettinelli, M.; Yang, H.; Huang, W.; Liu, X.
All-inorganic perovskite nanocrystal scintillators.
\textit{Nature} \textbf{2018}, 561, 88--93.

\bibitem{Protesescu2015} Protesescu, L.; Yakunin, S.; Bodnarchuk, M. I.; Krieg, F.; Caputo, R.; Hendon, C. H.; Yang, R. X.; Walsh, A.; Kovalenko, M. V.
Nanocrystals of cesium lead halide perovskites (CsPbX$_3$, X = Cl, Br, and I): Novel optoelectronic materials showing bright emission with wide color gamut.
\textit{Nano Lett.} \textbf{2015}, 15, 3692--3696.

\bibitem{Akkerman2018} Akkerman, Q. A.;  Rain\'o, G.;  Kovalenko, M. V.;  Manna, L.
Genesis, challenges and opportunities for colloidal lead halide perovskite nanocrystals.
\textit{Nature Mater.} \textbf{2018}, 17, 394.

\bibitem{Park2015} Park, Y.-S.; Guo, S.; Makarov, N. S.; Klimov, V. I.
Room temperature single-photon emission from individual perovskite quantum dots.
\textit{ACS Nano} \textbf{2015}, 9, 10386--10393.

\bibitem{Yu2021} Yu, B.; Zhang, C.; Chen, L.; Qin, Z.; Huang, X.; Wang X.; Xiao M.
Ultrafast dynamics of photoexcited carriers in perovskite semiconductor nanocrystals.
\textit{Nanophotonics} \textbf{2021}, 10, 1943--1965.

\bibitem{Li2017}
  Li, P.; Hu, C.; Zhou, L.; Jiang, J.; Cheng, Y.;, M. He; Liang, X.; Xiang, W.
Novel synthesis and optical characterization of CsPb$_2$Br$_5$ quantum dots in borosilicate glasses,
\textit{Mater. Lett.} \textbf{2017}, {209}, 483--485.

\bibitem{Liu2018}
   Liu, S.; Luo, Y.; He, M.; Liang, X.; Xiang, W.
Novel CsPbI$_3$ QDs glass with chemical stability and optical properties.
\textit{J. Eur. Ceram. Soc.} \textbf{2018}, {38}, 1998--2004.

\bibitem{Liu2018a}
  Liu, S. ; He, M.; Di, X.; Li, P.; Xiang, W.; Liang, X.
Precipitation and tunable emission of cesium lead halide perovskites (CsPbX$_3$, X = Br, I) QDs in borosilicate glass.
\textit{Ceram. Int.} \textbf{2018}, {44}, 4496--4499.

\bibitem{Ye2019}
   Ye, Y.; Zhang, W.; Zhao, Z.; Wang, J.; Liu, C.; Deng, Z.; Zhao, X.; Han, J.
Highly luminescent cesium lead halide perovskite nanocrystals stabilized in glasses for light-emitting applications.
\textit{Adv. Opt. Mater.} \textbf{2019}, {7}, 1801663.

\bibitem{kolobkova2021}
   Kolobkova, E. V.; Kuznetsova, M. S.; Nikonorov,  N. V.
   Perovskite CsPbX\textsubscript{3} (X = Cl, Br, I) nanocrystals in fluorophosphate glasses.
\textit{J. Non-Crystalline Solids} \textbf{2021}, {563}, 120811.

\bibitem{Belykh2022}  Belykh, V. V.;  Skorikov,  M. L.; Kulebyakina, E. V.; Kolobkova,  E. V.;  Kuznetsova, M. S.; Glazov, M. M.; Yakovlev, D. R.
Submillisecond spin relaxation in CsPb(Cl,Br)$_3$ perovskite nanocrystals in a glass matrix.
\textit{Nano Lett.}  \textbf{2022}, {22}, 4583--4588.

\bibitem{ekimov1981} Ekimov A. I.; Onushchenko, A. A. 
  Quantum size effect in three-dimensional microscopic semiconductor crystals.
  \textit{Pis'ma Zh. Eksp. Teor. Fiz.} \textbf{1981}, 34, 363 [JETP Lett. 34, 345 (1981)].

\bibitem{efros1982}  Efros, Al. L.; Efros, A. L. 
  Interband absorption of light in a semiconductor sphere.
  \textit{Fiz. Tekh. Poluprovodn.} \textbf{1982}, 16, 1209 [Sov. Phys. Semicond. 16, 772 (1982)].

\bibitem{ekimov1985}  Ekimov, A. I.; Efros, Al. L.; Onushenko, A. A.
  Quantum size effect in semiconductor microcrystals.
  \textit{Solid State Commun.} \textbf{1985}, 56, 921.

\bibitem{bawendi1993}  Murray, C. B.; Norris, D. J.; Bawendi, M. G.
  Synthesis and characterization of nearly monodisperse CdE (E = sulfur, selenium, tellurium) semiconductor nanocrystallites. 
  \textit{J. Am. Chem. Soc.} \textbf{1993}, 115, 19, 8706–8715. 

\bibitem{Talapin2010} 
   Talapin, D. V.; Lee, J.-S.; Kovalenko, M. V.; Shevchenko, E. V.
   Prospects of colloidal nanocrystals for electronic and optoelectronic applications.
\textit{Chemical Reviews} \textbf{2010}, 110, 389-458.

\bibitem{marzin1994}  Marzin, J. -Y.; G\'erard, J. -M.; Izra\"el, A.; Barrier, D.; Bastard, G. 
  Photoluminescence of single InAs quantum dots obtained by self-organized growth on GaAs. 
  \textit{Phys. Rev. Lett.} \textbf{1994}, 73, 716.

\bibitem{alferov1995}  Grundmann, M.; Christen, J.; Ledentsov, N. N.; B\"ohrer, J.; Bimberg, D.; Ruvimov, S. S.; Werner, P.; Richter, U.; G\"osele, U.; Heydenreich, J.; Ustinov, V. M.; Egorov, A. Yu.; Zhukov, A. E.; Kop'ev, P. S.; Alferov, Zh. I.
  Ultranarrow luminescence lines from single quantum dots. 
  \textit{Phys. Rev. Lett.} \textbf{1995}, 74, 4043.

\bibitem{lamb1881}  Lamb, H. On the vibrations of an elastic sphere. \textit{Proc. London Math. Soc.} \textbf{1881}, s1-13, 189.
 
\bibitem{saviot2011}  Saviot, L.; Mermet, A.; Duval, E. 
  Acoustic vibrations in nanoparticles. In \textit{Nanoparticles and Quantum Dots, Handbook of Nanophysics Vol. 3},  Sattler, K. D. Ed.; CRC Press, Boca Raton, \textbf{2011}; Chap. 11, pp. 11-1–11-16.

\bibitem{goupalov2020} Goupalov, S.V. 
  Classical problems in the theory of elasticity and the quantum theory of angular momentum. 
  \textit{Usp. Fiz. Nauk} \textbf{2020}, 190, 63 [Phys.-Usp. 63, 57 (2020)].

\bibitem{duval1992} Duval, E.
Far-infrared and Raman vibrational transitions of a solid sphere: Selection rules.
\textit{Phys. Rev. B.} \textbf{1992}, 46, {5795}.

\bibitem{goupalov1999} Gupalov, S. V.; Merkulov, I. A.
Theory of Raman light scattering by nanocrystal acoustic vibrations.
\textit{Phys. Solid State} \textbf{1999}, 41, {1349--1358}.

\bibitem{goupalov2000} Goupalov, S. V.; Ekimov, A. I.; Lublinskaya, O. G.; Merkulov, I. A. \textit{Raman Spectroscopy of Exciton-Acoustic Phonon Coupling in Semiconductor Nanocrystals.} In: Sadowski, M. L.; Potemski, M.; Grynberg, M. (eds) Optical Properties of Semiconductor Nanostructures. NATO Science Series, vol 81. Springer, Dordrecht. \textbf{2000}.

\bibitem{nestoklon2022} Nestoklon, M. O.; Goupalov, S. V.
Exciton interaction with acoustic phonons in PbS nanocrystals.
\textit{Phys. Rev. B} \textbf{2022}, 106, {045306}. 

\bibitem{saviot2009} Saviot, L.; Murray, D. B.
Acoustic vibrations of anisotropic nanoparticles.
\textit{Phys. Rev. B} \textbf{2009}, 79, {214101--214112}.

\bibitem{ng2022}  Ng, R. C.; Sachat, A. E.; Cespedes, F.; Poblet, M.; Madiot, G.; Jaramillo-Fernandez, J.; Florez, O.; Xiao, P.; Sledzinska, M.; Sotomayor-Torres, C. M.; Chavez-Angel, E.
  Excitation and detection of acoustic phonons in nanoscale systems. 
  \textit{Nanoscale} \textbf{2022}, 14, 13428.

\bibitem{fujii1996}  Fujii, M.; Kanzawa, Y.; Hayashi, S.; Yamamoto, K.
  Raman scattering from acoustic phonons confined in Si nanocrystals. 
  \textit{Phys. Rev. B} \textbf{1996}, 54, R8373.

\bibitem{saviot1996} Saviot, L.; Champagnon, B.; Duval, E.; Kudriavtsev, I. A.;  Ekimov, A. I.
Size dependence of acoustic and optical vibrational modes of CdSe nanocrystals in glasses.
\textit{J. Non-Crystalline Solids} \textbf{1996}, 197, {238--246}.

\bibitem{sirenko1998} Sirenko, A. A.; Belitsky, V. I.; Ruf, T.; Cardona, M.; Ekimov, A. I.;  Trallero-Giner, C.
Spin-flip and acoustic-phonon Raman scattering in CdS nanocrystals.
\textit{Phys. Rev. B} \textbf{1998}, 58, {2077--2087}.

\bibitem{saviot1998} Saviot, L.; Champagnon, B.; Duval, E.;  Ekimov, A. I.
Size-selective resonant Raman scattering in CdS doped glasses.
\textit{Phys. Rev. B} \textbf{1998}, 57, {341--346}.

\bibitem{Krauss1997}  Krauss, T. D.; Wise, F. W. 
  Coherent acoustic phonons in a semiconductor quantum dot. 
  \textit{Phys. Rev. Lett.} \textbf{1997}, 79, 5102.

\bibitem{Ikezawa2001}  Ikezawa, M.; Okuno, T.; Masumoto, Y.; Lipovskii, A. A.
  Complementary detection of confined acoustic phonons in quantum dots by coherent phonon measurement and Raman scattering. 
  \textit{Phys. Rev. B} \textbf{2001}, 64, 201315(R).

\bibitem{vlk2022}  Vlk, A.; Ledinsky, M.; Shiryaev, A.; Ekimov, E.; Stehlik, S.
  Nanodiamond size from low-frequency acoustic Raman modes. 
  \textit{J. Phys. Chem. C} \textbf{2022}, 126, 6318--6324.

\bibitem{Miyata2017}  Miyata, K.; Meggiolaro, D.; Trinh, M. T.; Joshi, P. P.; Mosconi, E.; Jones, S. C.; De Angelis, F.; Zhu, X.-Y. 
  Large polarons in lead halide perovskites.
  \textit{Sci. Adv.} \textbf{2017}, 3, e1701217. 

\bibitem{Ponce2019}  Ponc{\'e}, S.; Schlipf, M.; Giustino, F.
  Origin of low carrier mobilities in halide perovskites. 
  \textit{ACS Energy Lett.} \textbf{2019}, 4, 456.

\bibitem{Fu2017} Fu, M.; Tamarat, P.; Huang, H.; Even, J.; Rogach, A. L.; Lounis, B.
Neutral and charged exciton fine structure in single lead halide perovskite nanocrystals revealed by magneto-optical spectroscopy.
\textit{Nano Lett.} \textbf{2017}, 17, 2895--2901.

\bibitem{Cho2021}  Cho, K.; Yamada, T.; Tahara, H.; Tadano, T.; Suzuura, H.; Saruyama, M.; Sato, R.; Teranishi, T.; Kanemitsu, Y.
  Luminescence fine structures in single lead halide perovskite nanocrystals: Size dependence of the exciton--phonon coupling. 
  \textit{Nano Lett.} \textbf{2021}, 21, 7206.

\bibitem{Iary2021}  Iaru, C. M.; Brodu, A.; van Hoof, N. J. J.; ter Huurne, S. E. T.; Buhot, J.; Montanarella, F.; Buhbut, S.; Christianen, P. C. M.; Vanmaekelbergh, D.; de Mello Donega, C.; Rivas, J. G.; Koenraad, P. M.; Silov, A. Y.
  Fr{\"o}hlich interaction dominated by a single phonon mode in CsPbBr\textsubscript{3}. 
  \textit{Nature Commun.} \textbf{2021}, 12, 5844.

\bibitem{Amara2023}   Amara, M.-R.; Said, Z.; Huo, C.; Pierret, A.; Voisin, C.; Gao, W.; Xiong, Q.; Diederichs, C.
Spectral fingerprint of quantum confinement in single CsPbBr$_3$ nanocrystals. 
\textit{Nano Lett.} \textbf{2023}, 23, 3607.

\bibitem{Lv2021}  Lv, Y.; Yin, C.; Zhang, C.; Wang, X.; Yu, Z.-G.; Xiao, M.
Exciton-acoustic phonon coupling revealed by resonant excitation of single perovskite nanocrystals. 
\textit{Nature Commun.} \textbf{2021}, 12, 2192.

\bibitem{meliakov2025} Meliakov, S. R.; Zhukov, E. A.; Belykh; V. V., Nestoklon; M. O., Kolobkova, E. V.; Kuznetsova, M. S.; Bayer, M.;  Yakovlev, D. R.
  Land\'e $g$-factors of electrons and holes strongly confined in CsPbI$_3$ perovskite nanocrystals in glass. 
  \textit{Nanoscale}  \textbf{2025}, 17, 6522.

\bibitem{nestoklon2023} Nestoklon, M. O.; Kirstein, E.; Yakovlev, D. R.; Zhukov, E. A.; Glazov, M. M.; Semina, M. A.; Ivchenko, E. L.;  Kolobkova, E. V.; Kuznetsova, M. S.; Bayer, M.
  Tailoring the electron and hole Land\'e factors in lead halide perovskite nanocrystals by quantum confinement and halide exchange.
  \textit{Nano Lett.} \textbf{2023}, 23, 8218–8224.

\bibitem{meliakov2024} Meliakov, S. R.; Zhukov, E. A.; Belykh, V. V.; Nestoklon, M. O.; Kolobkova, E. V.; Kuznetsova, M. S.; Bayer, M.; Yakovlev, D. R.
  Temperature dependence of the electron and hole Land\'e $g$-factors in CsPbI$_3$ nanocrystals embedded in a glass matrix. \textit{Nanoscale} \textbf{2024}, 16, {21496}.

\bibitem{bataev2024}  Bataev, M. N.; Kuznetsova, M. S.; Pankin, D. V.; Smirnov, M. B.; Verbin, S. Yu.; Ignatiev, I. V.; Eliseyev, I. A.; Davydov, V. Yu.; Smirnov, A. N.; Kolobkova, E. V.  
  Electron-phonon interaction in perovskite nanocrystals in fluorophosphate glass matrix.
  \textit{Semiconductors} \textbf{2024}, 58, {103-109}.

\bibitem{Martin1971} Martin, R. M. 
  Theory of the one-phonon resonance Raman effect.
  \textit{Phys. Rev. B} \textbf{1971}, 4, 3676.

\bibitem{saviot2021} Saviot, L.
Free vibrations of anisotropic nano-objects with rounded or sharp corners.
\textit{Nanomaterials} \textbf{2021}, 11, {1838}.

\bibitem{Perdew2008}  Perdew, J. P.; Ruzsinszky, A.; Csonka, G. I.; Vydrov, O. A.; Scuseria, G. E.; Constantin, L. A.; Zhou, X.; Burke, K.
Restoring the density-gradient expansion for exchange in solids and surfaces.
\textit{Phys. Rev. Lett.} \textbf{2008}, 100, 136406.

\bibitem{Blaha2020} Blaha, P.; Schwarz, K.; Tran, F.; Laskowski, R.; Madsen, G. K. H.; Marks, L. D.
WIEN2k: An APW+lo program for calculating the properties of solids. 
\textit{J. Chem. Phys.} \textbf{2020}, 152, {074101}.

\bibitem{Anderson1963} Anderson, O. L.
A simplified method for calculating the debye temperature from elastic constants.
\textit{J. Phys. Chem. Solids} \textbf{1963}, 24, {909--917}.

\bibitem{Wherrett1986}
Wherrett, B. S.
Group Theory for Atoms, Molecules and Solids; Prentice Hall International: Englewood Cliffs, New Jersey, \textbf{1986}.

\bibitem{Nestoklon2018} Nestoklon, M. O.;  Goupalov, S. V.;  Dzhioev, R. I.; Ken, O. S.; Korenev, V. L.; Kusrayev, Yu. G.; Sapega, V. F.; de Weerd, C.; Gomez, L.; Gregorkiewicz, T.; Lin, J.; Suenaga, K.; Fujiwara, Y.; Matyushkin, L. B.; Yassievich, I. N.
Optical orientation and alignment of excitons in ensembles of inorganic perovskite nanocrystals.
\textit{Phys. Rev. B} \textbf{2018}, {97}, 235304.

\bibitem{Tamarat2022}  Tamarat, P.; Prin, E.; Berezovska, Y.; Moskalenko, A.; Nguyen, T. P. T.; Xia, C.; Hou, L.; Trebbia, J.-B.; Zacharias, M.; Pedesseau, L.; Katan, C.; Bodnarchuk, M. I.; Kovalenko, M. V.; Even, J.; Lounis, B.
Universal scaling laws for charge-carrier interactions with quantum confinement in lead-halide perovskites.
\textit{Nature Commun.} \textbf{2023}, {14}, 229.

\bibitem{Han2022}  Han, Y.; Liang, W.; Lin, X.; Li, Y.; Sun, F.; Zhang, F.; Sercel, P. C.; Wu, K.
  Lattice distortion inducing exciton splitting and coherent quantum beating in CsPbI$_3$ perovskite quantum dots.
  \textit{Nature Materials} \textbf{2022}, {21}, 1282.

\end{thebibliography}

\clearpage

\begin{thebibliography}{99}

\bibitem{IvchenkoBooksi} Ivchenko, E. L. 
	\textit{Optical Spectroscopy of Semiconductor Nanostructures}; Alpha Science, \textbf{2005}

\bibitem{meliakov2025si} Meliakov, S. R.; Zhukov, E. A.; Belykh; V. V., Nestoklon; M. O., Kolobkova, E. V.; Kuznetsova, M. S.; Bayer, M.;  Yakovlev, D. R.
  Land\'e $g$-factors of electrons and holes strongly confined in CsPbI$_3$ perovskite nanocrystals in glass. 
  \textit{Nanoscale}  \textbf{2025}, 17, 6522.


\bibitem{Blaha2020si} Blaha, P.; Schwarz, K.; Tran, F.; Laskowski, R.; Madsen, G. K. H.; Marks, L. D.
WIEN2k: An APW+lo program for calculating the properties of solids. 
\textit{J. Chem. Phys.} \textbf{2020}, {152}, {074101}.

\bibitem{Jamal2018si} Jamal, M.; Bilal, M.; Ahmad, I.; Jalali-Asadabadi, S.
IRelast package. \textit{J. Alloys and Compounds} \textbf{2018}, {735}, {569}.

\bibitem{LLIVsi}  Berestetskii, V. B.; Pitaevskii, L. P.; Lifshits, E. M.
\textit{Quantum Electrodynamics: Volume 4}; Butterworth-Heinemann, \textbf{1982}.

\bibitem{duval1992si} Duval, E.
Far-infrared and Raman vibrational transitions of a solid sphere: Selection rules.
\textit{Phys. Rev. B.} \textbf{1992}, {46}, {5795}.

\bibitem{gupalov2006si} Goupalov, S. V.; Saviot, L.; Duval, E.
Comment on “Infrared and Raman selection rules for elastic vibrations of spherical nanoparticles”.
\textit{Phys. Rev. B.} \textbf{2006}, {74}, {197401}.

\bibitem{goupalov1999si} Gupalov, S. V.; Merkulov, I. A.
Theory of Raman light scattering by nanocrystal acoustic vibrations.
\textit{Phys. Solid State} \textbf{1999}, {41}, {1349--1358}.

\bibitem{Wherrett1986si}
Wherrett, B. S.
Group Theory for Atoms, Molecules and Solids; Prentice Hall International: Englewood Cliffs, New Jersey, \textbf{1986}.


\bibitem{nestoklon2023si} Nestoklon, M. O.; Kirstein, E.; Yakovlev, D. R.; Zhukov, E. A.; Glazov, M. M.; Semina, M. A.; Ivchenko, E. L.;  Kolobkova, E. V.; Kuznetsova, M. S.; Bayer, M.
Tailoring the electron and hole Land\'e factors in lead halide perovskite nanocrystals by quantum confinement and halide exchange, 
\textit{Nano Lett.} \textbf{2023}, {23}, 8218--8224.

\bibitem{galkowski2016si} 
Galkowski, K.; Mitioglu, A.; Miyata, A.; Plochocka, P.; Portugall, O.; Eperon, G. E.; Wang, J. T.-W.; Stergiopoulos, T.; Stranks, S. D.; Snaith, H. J.; Nicholas, R. J.
Determination of the exciton binding energy and effective masses for methylammonium and formamidinium lead tri-halide perovskite semiconductors.
\textit{Energy Environ. Sci.} \textbf{2016}, {9}, {962-970}.
	



\end{thebibliography}
\end{document}